\documentclass[10pt,twoside,a4paper,reqno]{amsart}
\usepackage{amsfonts,amsthm,latexsym,amsmath,amssymb,amscd,epsf}
\usepackage{epsfig}
\usepackage[dvips]{color}
\definecolor{LightGreen1}{rgb}{0.2,0.8,0.2}
\definecolor{DarkGray}{rgb}{0.4,0.4,0.4}
\definecolor{GhostWhite}{rgb}{0.97,0.97,1.}

\theoremstyle{plain}
\newtheorem{theo}           {Theorem}
\newtheorem{pro}            {Proposition}
\newtheorem{coro}           {Corollary}
\newtheorem{lemm}           {Lemma}
\newtheorem{conj}           {Conjecture}

\theoremstyle{definition}
\newtheorem{pr}              {Problem}
\newtheorem*{ack}            {Acknowledgements}

\theoremstyle{remark}
\newtheorem{rem}             {Remark}

\theoremstyle{definition}

\newenvironment{theorem}{\begin{theo}}{\end{theo}}
\newenvironment{proposition}{\begin{pro}}{\end{pro}}

\newenvironment{lemma}{\begin{lemm}}{\end{lemm}}
\newenvironment{remark}{\begin{rem}}{\end{rem}}

\newenvironment{problem}{\begin{pr}}{\end{pr}}
\newenvironment{conjecture}{\begin{conj}}{\end{conj}}

\newcommand \C {\mathcal C}

\newcommand \cc {\mathfrak c}

\newcommand \ga {\gamma}

\newcommand \bR {\mathbb R}
\newcommand \bC {\mathbb C}
\newcommand \bZ {\mathbb Z}
\newcommand\al {\alpha}
\newcommand \be {\beta}

\newcommand{\Com}{\mathbb{C}}

\newcommand{\iu}{\mathrm{i}}
\newcommand{\V}{\mathfrak V}

\begin{document}

\title[On spectral polynomials of the Heun equation. I]
{On spectral polynomials of the Heun equation. I.}

\author[B.~Shapiro]{Boris Shapiro}
\address{Department of Mathematics, Stockholm University, SE-106 91
Stockholm,
      Sweden}
\email{shapiro@math.su.se}

\author[M.~Tater]{Milo\v{s} Tater}
\address{Department of Theoretical Physics, Nuclear Physics Institute, 
Academy of Sciences, 250\,68 \v{R}e\v{z} near Prague, Czech
Republic}
\email{tater@ujf.cas.cz}


\begin{abstract}
The {\em classical Heun equation}  has the form
$$
         \left\{Q(z)\frac
         {d^2}{dz^2}+P(z)\frac{d}{dz}+V(z)\right\}S(z)=0,
   $$
where $Q(z)$ is a cubic complex polynomial,  $P(z)$ is
a  polynomial of degree at most $2$ and $V(z)$ is at most linear. 
In the second half of the nineteenth century E.~Heine and  T.~Stieltjes in \cite{He}, \cite{St} initiated the study of the set of all  $V(z)$ for which the above equation has a polynomial solution $S(z)$ of a given degree $n$. The main goal of  the present paper is to study the union of the roots of  the latter set of $V(z)$'s when $n\to\infty$. We formulate an  intriguing conjecture of K.~Takemura describing the limiting set and give a substantial amount of  additional information obtained using some technique developed in \cite{KvA}. 
\end{abstract}

\maketitle

\section{Introduction and Main Results}

A {\em generalized Lam\'e equation} is a second
order differential equation of the form
\begin{equation}\label{eq:comLame}
         \left\{Q(z)\frac
         {d^2}{dz^2}+P(z)\frac{d}{dz}+V(z)\right\}S(z)=0,
         \end{equation}
where $Q(z)$ is a complex polynomial of degree $l$ and $P(z)$ is
a complex polynomial of degree at most $l-1$, see  \cite{WW}.  
It was first shown by Heine \cite{He} that if the coefficients of $Q(z)$ and $P(z)$ are algebraically independent, i.e. do not satisfy any algebraic equation with integer coefficients then  
for an arbitrary positive integer $n$ there are exactly
$\binom{n+l-2}{n}$ polynomials $V(z)$ such that  \eqref{eq:comLame} has a solution $S(z)$ which is a polynomial of degree $n$. As was recently shown in \cite{Sh} for any equation  \eqref{eq:comLame} with $\deg Q(z)=l, \deg P(z)\le l-1$,  and any positive $n$ the set $\V_n$ of all $V(z)$ giving a polynomial solution $S(z)$ of degree $n$ is always finite and its cardinality is at most $\binom{n+l-2}{n}$. Below we concentrate on the  classical case $l=\deg Q(z)=3$ which is better known under the name {\em the Heun differential equation}, see e.g. \cite {Heun} and study the union of all  roots of polynomials $V(z)$ belonging to $\V_n$ as $n\to\infty$. Note that if $l=\deg  Q(z)=3$ then $V(z)$ is at most linear and
that for a given value of the positive integer $n$ there are at most $n+1$ such polynomials.

No essential results in this direction seems to be known. One of the few exceptions is a classical 
proposition of  P\'olya, \cite{Po} claiming that if the rational function $\frac{P(z)}{Q(z)}$ has all positive residues then any root of any $V(z)$ as above and of any $S(z)$ as above lie within   $Conv_Q$ where $Conv_Q$ is the convex hull of the set of all roots of $Q(z)$. 

Before we move further let us formulate appropriate versions of two main results of \cite{Sh} generalizing the above statements  of  Heine and 
P\'olya. 
 
 \begin{theorem}\label{th:my} For any polynomial $Q(z)$ of degree $l$ and any polynomial $P(z)$ of degree at most $l-1$ 
 \begin{itemize}
 \item there exists $N$ such that for any $n\ge N$ there exist exactly $\binom{n+l-2}{n}$ polynomials $V(z), \deg V(z)=l-2$ counted with appropriate multiplicity such that \eqref{eq:comLame} has a polynomial solution $S(z)$ of degree exactly $n$;
 \item for any $\epsilon >0$ there exists $N_\epsilon$ such that for  any $n\ge N_\epsilon$ any root of any above $V(z)$ and $S(z)$ lie in the $\epsilon$-neighborhood of $Conv_Q$.
 \end{itemize}
  \end{theorem}
  
  Applying the latter result to the  situation $l=3$, i.e to the Heun equation we can introduce the set $\mathcal V_n$ consisting of polynomials $V(z)$ giving a polynomial solution $S(z)$ of \eqref{eq:comLame} of degree $n$; each such $V(z)$ appearing the number of times equal to its multiplicity. Then by the above results  the set $\mathcal V_n$ will contain exactly $n+1$ linear polynomials for all sufficiently large $n$. It will be convenient to introduce a  sequence $\{Sp_n(\lambda)\}$ of {\em spectral polynomials} where the $n$-th spectral polynomial is defined by 
$$Sp_n(\lambda)=\prod_{j=1}^{n+1}(\lambda-t_{n,j}),$$
where $t_{n,j}$ is the unique root of the $j$-th polynomial in $\mathcal V_n$  in any fixed ordering.  ($Sp_n(\lambda)$ will be well-defined for all sufficiently large  $n$.)

Associate to $Sp_n(\lambda)$ the finite measure
$$\mu_n=\frac{1}{n+1}\sum_{j=1}^{n+1}{\delta(z-t_{n,j})},$$
where $\delta(z-a)$ is the Dirac measure supported at $a$.
The measure $\mu_n$  obtained in this way is clearly a real probability measure which 
one usually refers to as the {\em root-counting measure} of the polynomial $Sp_n(\lambda)$. 

The starting point of this project was some numerical results for the distribution of roots of $Sp_n(\lambda)$ obtained by the first author about 5 years ago and illustrated on the next figure. 

\begin{figure}[!htb]
\centerline{\hbox{\epsfysize=4cm\epsfbox{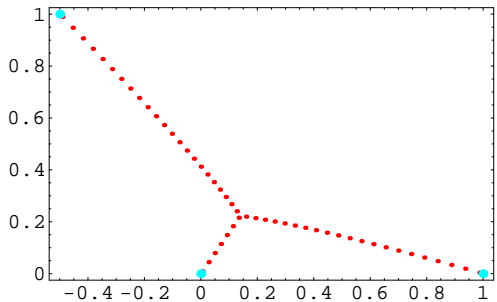}}}
\caption{The roots of the spectral polynomial $Sp_{50}(\lambda)$ for the classical Lam\'e equation 
$\left\{Q(z)\frac
         {d^2}{dz^2}+\frac{1}{2}Q'(z)\frac{d}{dz}+V(z)\right\}S(z)=0$, with 
$Q(z)=z(z-1)\left(z+\frac{1}{2}-i\right)$.}
\label{fig1}
\end{figure}

Extensive numerical experiments strongly suggest that the following holds. 

\begin{conjecture}[Shapiro-Tater]\label{sh-ta1}
For any equation \eqref{eq:comLame} the sequence $\{\mu_n\}$ of the root-counting measures of its spectral polynomials converges  to a probability measure $\mu$ supported on the union of three curved segments located inside $Conv_Q$ and connecting the three roots of $Q(z)$ with a certain interior point, see Fig.~\ref{fig1}. Moreover, the limiting measure $\mu$ depends only on $Q(z)$, i.e. is {\em independent} of $P(z)$. 
\end{conjecture}

An elegant description of the support of $\mu$ was suggested to us by Professor K.~Takemura, \cite{TaM}. 

Denote the three roots of $Q(z)$ by $a_1,a_2,a_3$. For $i\in\{1,2,3\}$  consider the curve $\gamma_i$ given as the set of all $b$ satisfying the relation:  
\begin{equation}\label{TK}
\int_{a_j}^{a_k}\sqrt{\frac{b-t}{(t-a_1)(t-a_2)(t-a_3)}}dt\in \bR,
\end{equation}
here $j$ and $k$ are the remaining two indices in $\{1,2,3\}$ in any order and the integration is taken over the straight interval connecting $a_j$ and $a_k$. One can see that $a_i$ belong to $\gamma_i$ and that these three curves connect the corresponding $a_i$ with a common point within $Conv_Q$. Take a segment of $\gamma_i$ connecting $a_i$ with the common intersection point of all $\gamma$'s. Let us denote the union of these three segments by $\Gamma_Q$. 

\begin{conjecture}[Takemura] \label{takemura} The support of the limiting root-counting measure $\mu$ coincides with the above $\Gamma_Q$. 
\end{conjecture} 

\medskip
The above description of $\Gamma_Q$  led us to the following reformulation  of Takemura's conjecture. 

\begin{proposition}
 \label{sh-ta2} The above set  $\Gamma_Q$ coincides with the continuum of minimal logarithmic capacity connecting the roots of $Q(z)$.
\end{proposition}

Notice that  Goluzin's classical problem of finding  the continuum of minimal capacity connecting a given $n$-tuple of points in $\bC$ was completely solved for $n=3$ by G.~Kuzmina in \cite{Ku1}, see also \cite{Ku2}. 

In  the joint with Professor Takemura follow-up of the present paper \cite{STT} we will  completely  settle the above Conjecture~\ref{takemura} and Proposition~\ref{sh-ta2} using some methods and results presented below.  In the present paper generalizing the technique of \cite {KvA}  we study a different probability measure which is easily described and from which the measure  $\mu$ (if it exists)  is obtained by the inverse  balayage, i.e. the support of $\mu$ will be contained in the support of the measure which we construct and they have the same logarithmic potential outside the support of the latter  one. This measure will be uniquely determined by  the choice of a root of $Q(z)$ and thus we are in fact constructing three different measures having the same measure $\mu$ as their inverse balayage.    

\subsection{Constructing the measure} 
Choosing one of the three vertices $a_i,\; i=\{1,2,3\}$ consider the unique ellipse $E_i$ which: a) passes through $a_i$ and b) has $a_j, a_k$ as its foci. The constructed probability measure $M_i$ is supported on the elliptic domain $\tilde E_i$ bounded by $E_i$. We need the following notion. 

Given two distinct points $\alpha_1\neq \alpha_2$ on $\bC$ define the {\em arcsine measure}  
$\omega_{[\alpha_1, \alpha_2]}$ of the interval $[\alpha_1,\alpha_2]$ as  the measure supported on $[\alpha_1,\alpha_2]$ and whose density at a point $t\in [\alpha_1,\alpha_2]$ equals $\frac{1}{\pi \sqrt{|(t-\alpha_1)(t-\alpha_2)|}}$.

\begin{figure}[!htb]
\centerline{\hbox{\epsfysize=4.5cm\epsfbox{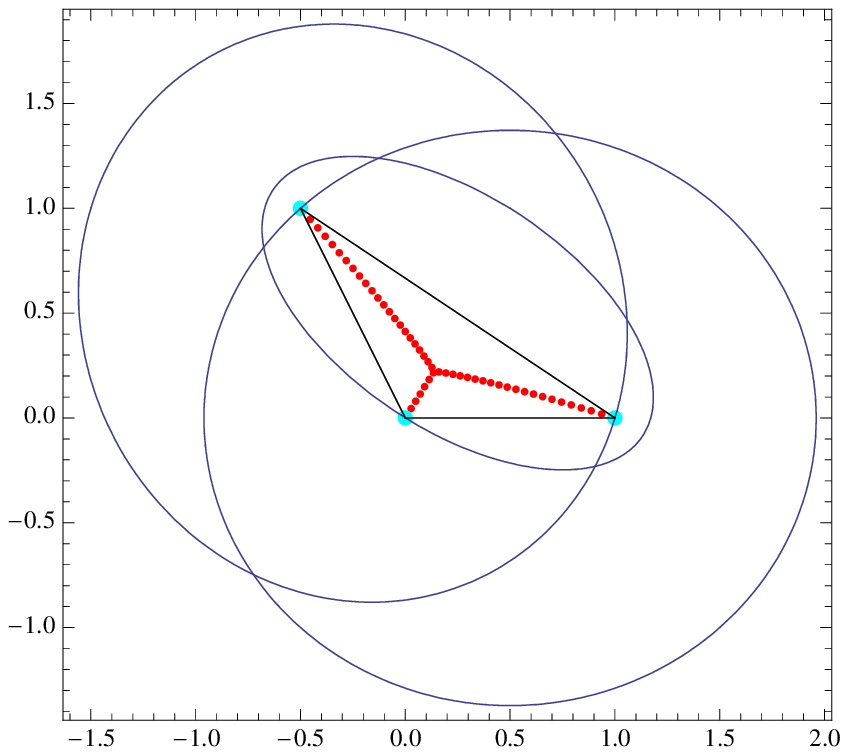}}{\hbox{\epsfysize=4.5cm\epsfbox{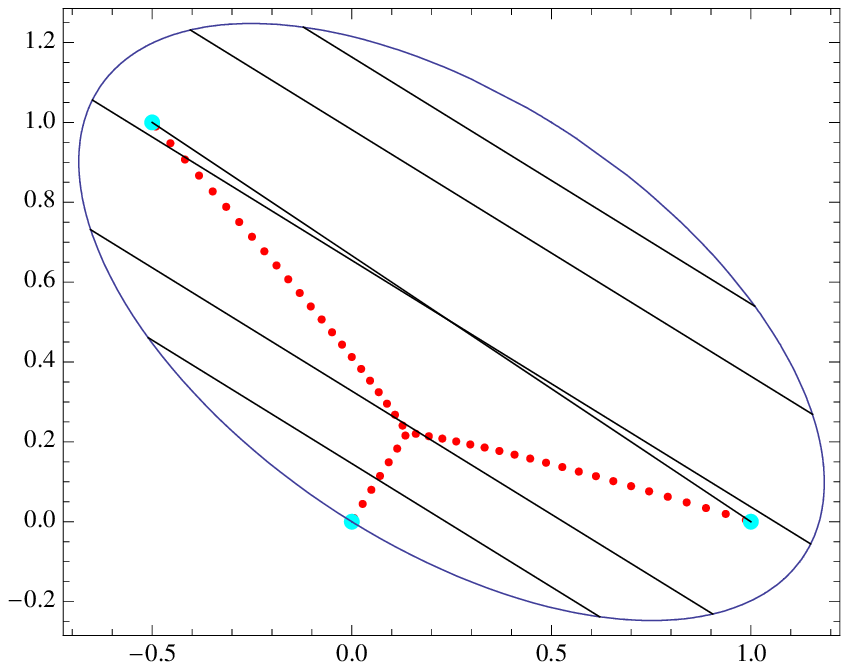}}}}
\caption{a) The measure $\mu$ and the three ellipses $E_1,E_2,E_3$ for $Q(z)=z(z-1)\left(z+\frac{1}{2}-i\right)$. b) The measure $\mu$, ellipse $E_1$, and several straight segments belonging to the family $\Phi_1$.}
\label{fig2}
\end{figure}

  To describe  the measure $M_i$ consider the family of straight lines parallel to the tangent line to the ellipse $E_i$ at $a_i$. Take now the family $\Phi_i$ of  intervals obtained by intersection of the latter straight lines with the elliptic domain $\tilde E_i$. Denote by $-v_i$ the vector connecting $a_i$ with its opposite point on $E_i$, i.e. draw the straight line through $a_i$ and the center of $E_i$ till it hits $E_i$ again and take the difference of the latter and the former points. (One can easily check that if we introduce a new variable $z_i=z-a_i$ and express $Q(z)=z_i^3+v_iz^2_i+w_iz_i$ then the above vector will be exactly $-v_i$ in the expression for $Q(z)$ which explains our notation.) Now parameterize the above family 
  $\Phi_i$ of the intervals by their middle points using the formula
   $-v_i\theta^2, \; \theta\in[0,1].$
    Consider the family  $\mu_\theta$ of arcsine measures of these intervals. Finally the required measure $M_i$ is obtained by the averaging of $\mu_\theta$ w.r.t. parameter $\theta$, i.e. $M_i=\int_0^1\mu_\theta d\theta$, see Fig.~\ref{fig2}b).

\medskip
Now we can finally formulate the main results of this paper.  

\begin{theorem}\label{th:main} If the measure $\mu$ in Conjecture~\ref{sh-ta1} exists then each of the measures $M_i,\; i\in\{1,2,3\}$ have $\mu$ as its inverse balayage, i.e. $\mu$ and $M_i$ have the same 
logarithmic potential  (or the same Cauchy transform) outside the ellipse $E_i$  and the support of $\mu$ is contained inside the support of $M_i$.
\end{theorem}

By definition  the Cauchy transform $\C_\nu(z)$ and the logarithmic potential $pot_\nu(z)$ of a (complex-valued) measure $\nu$ supported in $\bC$ are given by: 
$$\C_\nu(z)=\int_{\bC}\frac{d\nu(\xi)}{z-\xi}\quad\text{    and    }\quad pot_\nu(z)=\int_{\bC}\log|z-\xi|{d\nu(\xi)}.$$
About the properties of the Cauchy trasform and the logarithmic potential of a measure consult e.g. \cite {Ga}. 

\begin{rem} Theorem~\ref{th:main} is so far a conditional statement. For technical reasons complete proofs of the existence, uniqueness and several other properties of $\mu$ are postponed until \cite {STT}. 

\end{rem}

Denote by $\C_{Q_i}(z)$ the Cauchy transform of the measure $M_i,\;i=1,2,3$. The next result shows that each Cauchy transform $\C_{Q_i}(z)$ satisfies outside the elliptic domain $\tilde{E}_i$ the following nice linear non-homogeneous second order differential equation (similar to the one obtained earlier in \cite{BSh}).

\begin{theorem}\label{th:eq}
The Cauchy transforms $\C_{Q_i}(z)$ of the measures $M_i,\;i=1,2,3$ defined in
Theorem~\ref{th:main} satisfy outside the ellipses $E_i$ one and the same linear non-homogeneous differential equation: 
\begin{equation}\label{eq:Heun}
Q(z)\C''_{Q_i}(z)+Q'(z)\C'_{Q_i}(z)+\frac{Q''(z)}{8}\C_{Q_i}(z)+\frac{Q^{\prime\prime\prime}(z)}{24}=0.
\end{equation}
\end{theorem}

\begin{ack}
 We are very grateful to Professor K.~Takemura of 
Yokohama City University for a number of illuminating discussions prior,
during, and after  his visit to Stockholm in September 2007. We are obliged to Professor A.~Kuijlaars for clarification of his joint paper \cite{KvA} and to Professor A.~Mart\'inez-Finkelshtein for the interest in our work. We want to thank Professor G.~V.~Kuzmina for the patient explanation of her related results and Professor J.-E.~Bj\"ork for the help in manipulations with complicated integrals depending on parameters.  Research of the second author  was supported by the Czech Ministry of Education, Youth and Sports within the project LC06002.
\end{ack}

\section{Proof of Theorem~\ref{th:main}}

\begin{proof}[Proof of Theorem~\ref{th:main}]
It essentially follows from the stronger version of the main result of \cite{KvA} which we present below.  First we express the polynomial $Sp_n(\lambda)$ as the
characteristic polynomial of a certain matrix. In order to make this matrix
tridiagonal  we assume as above  that the root $a_i$ is placed at the origin. In order to simplify the notation we drop the index $i$ assuming that $z$ is already the appropriate coordinate. Set 
$$Q(z)=z^3+vz^2+wz.$$
 Consider the operator 
$$T=(z^3+vz^2+wz)\frac{d^2}{dz^2}+(\al z^2+\be z+\ga)\frac{d}{dz}
-\theta_n(z-\lambda),$$
where $v,w,\al,\be,\ga$ are fixed coefficients of $Q(z)$ and $P(z)$ respectively  and $\theta_n, \lambda$ are variables.
Assuming  that $S(z)=u_0z^n+u_1z^{n-1}+\ldots+u_n$ with undetermined
     coefficients  $u_i$, $0\le i\le n$, and in order to solve the Heine-Stieltjes problem described in the introduction we will be  looking for the values of
$\theta_n, \lambda$ and $u_i$, $0\le i\le n$, such that $T(S(z))=0$. Note that
     $T(S(z))$ is in general a polynomial of degree $n+1$ whose leading
coefficient equals $u_0[n(n-1)+\al n -\theta_n]$. To get a non-trivial
solution we therefore set
$$\theta_n=n(n-1+\al).$$
Straightforward computations show that the coefficients of
     the successive powers  $z^n, z^{n-1},\ldots, z^0$ in
     $T(S(z))$ can be expressed in the  form of a matrix product
$M_nU$, where $U=(u_0,u_1,\ldots,u_n)^T$ and  $M_n$ is the following
tridiagonal $(n+1)\times(n+1)$ matrix
$$
M_n:=\begin{pmatrix}
\lambda-\xi_{n,1}&\al_{n,2}&0&0&\cdots&0&0\\
\ga_{n,2}&\lambda-\xi_{n,2}&\al_{n,3}&0&\cdots&0&0\\
0&\ga_{n,3}&\lambda-\xi_{n,3}&\al_{n,4}&\cdots&0&0\\
\vdots&\vdots&\ddots&\ddots&\ddots&\vdots&\vdots\\
0&0&0&\ddots&\ddots&\al_{n,n}&0\\
0&0&0&\cdots&\ga_{n,n}&\lambda-\xi_{n,n}&\al_{n,n+1}\\
0&0&0&\cdots&0&\ga_{n,n+1}&\lambda-\xi_{n,n+1}
\end{pmatrix}
$$
with
\begin{equation}\label{eq:extra}
\begin{split}
\xi_{n,i}&=-\frac{v(n-i)(n-i+1)+\be (n-i+1)}{\theta_n},
\quad i\in\{1,\ldots,n+1\}, \\
\al_{n,i}&=\frac{(n-i)(n-i+1)+\al(n-i+1)}{\theta_n}-1,
\quad i\in\{2,\ldots,n+1\}, \\
\ga_{n,i}&=\frac{w(n-i+1)(n-i+2)+\ga (n-i+2)}{\theta_n},
\quad i\in\{2,\ldots,n+1\}.
\end{split}
\end{equation}

A similar matrix can be found in \cite{He} and also  in \cite{Tu}.
The matrix $M_n$ depends linearly on the indeterminate $\lambda$ which appears
only on its main diagonal. Obviously if the linear homogeneous system $M_nU=0$ is to have
a nontrivial solution $U=(u_0,u_1,...,u_n)^T$ the determinant
of $M_n$ has to vanish. This gives the required polynomial equation
$$Sp_n(\lambda)=\det(M_n)=0.$$

The sequence of polynomials $\{Sp_n(\lambda)\}_{n\in\bZ_+}$ does
not seem to satisfy any
reasonable recurrence relation. In order to overcome this difficulty and to
be able to use  the  technique of $3$-term recurrence relations  with
variable coefficients (which is applicable since $M_n$ is tridiagonal) we
extend the above polynomial sequence by introducing an additional parameter.
Namely, define
$$Sp_{n,i}(\lambda)=\det M_{n,i},\quad i\in\{1,\ldots,n+1\},$$
where $M_{n,i}$ is the upper $i\times i$ principal submatrix of $M_n$.
One can easily check (see, e.g., \cite[p.~20]{Ar})
that the following $3$-term relation holds
\begin{equation}\label{eq:3term}
Sp_{n,i}(\lambda)=(\lambda-\xi_{n,i}) Sp_{n,i-1}(\lambda) - \psi_{n,i} Sp_{n,i-2}(\lambda),
\quad i\in\{1,\ldots,n+1\},
\end{equation}
where $\xi_{n,i}$ is as in \eqref{eq:extra} and
\begin{equation}\label{eq:coeffs}
\psi_{n,i}=\al_{n,i}\ga_{n,i},\quad i\in\{2,\ldots,n+1\}.
\end{equation}
Here we use the (standard) initial conditions
$Sp_{n,0}(\lambda)=1$, $Sp_{n,-1}(\lambda)=0$.  It is well-known that if all
$\xi_{n,i}$'s are real and all $\psi_{n,i}$'s are  positive then the
polynomials $Sp_{n,i}(\lambda)$, $i\in\{0,\ldots, n+1\}$,
form a finite sequence of orthogonal polynomials.  In particular, all their
roots are real. 
 In our case however  these coefficients are complex. To complete the proof of
Theorem~\ref{th:main} we state the following generalization of  \cite[Theorem 1.4]{KvA} which translated in our notation
claims the following. 

\begin{theorem}[A.~Kuijlaars - W.~Van Assche]\label{KvA} If there exist two continuous functions $\xi(\tau)$ and
$\psi(\tau)$, $\tau\in [0,1]$, such that
$$\lim_{i/(n+1)\to \tau} \xi_{i,n}=\xi(\tau),\quad
\lim_{i/(n+1)\to \tau} \psi_{i,n}=\psi(\tau),\quad
\,\,\,\,\forall\tau\in [0,1],$$
then the asymptotic root-counting measure $\mu$ of the
polynomial sequence $\{Sp_n(\lambda)\}_{n\in\bZ_+}=\{Sp_{n,n+1}(\lambda)\}_{n\in\bZ_+}$ 
(if it exists) and the average $M$ of the acsine measures  
 given by
$$M=\int_0^1
\omega_{\left[\xi(\tau)-2\sqrt{\psi(\tau)},
\xi(\tau)+2\sqrt{\psi(\tau)}\right]}d\tau,$$
have the same logarithmic potential outside the union of their supports.
\end{theorem}

 Recall that for a pair of distinct complex number $\alpha_1\neq \alpha_2$ the  arcsine measure  
$\omega_{[\alpha_1, \alpha_2]}$ is  the measure supported on $[\alpha_1,\alpha_2]$ and whose density at a point $t\in [\alpha_1,\alpha_2]$ equals $\frac{1}{\pi \sqrt{|(t-\alpha_1)(t-\alpha_2)|}}$. 

\begin{remark} Although Theorem~\ref{KvA} is not explicitly stated in \cite{KvA} it is very similar and its proof is completely parallel to that of Theorem~1.4 from this paper. 
\end{remark}

\medskip
 From the explicit formulas for $\xi_{n,i}$ and $\psi_{n,i}$
(see \eqref{eq:extra} and \eqref{eq:coeffs}) one easily gets
\begin{equation*}
\begin{split}
\xi(\tau)&=\lim_{i/(n+1)\to \tau} \xi_{i,n}=-v(1-\tau)^2, \\
\psi(\tau)&=\lim_{i/(n+1)\to \tau} \psi_{i,n}=-w(1-(1-\tau)^2)(1-\tau)^2.
\end{split}
\end{equation*}
Notice that the above limits are independent of the coefficients
$\al,\be,\ga$ of the polynomial $P(z)$. \end{proof}

\begin{lemma}\label{ellipse}
The parametric curve $\Gamma$ given in the above notation  by the formula  $\xi(\tau)\pm 2\sqrt{\psi(\tau)},\;\tau\in [0,1]$ is the ellipse passing through the origin and given in coordinates $x=Re(z), y=Im(z)$ by the equation 
\begin{equation}\label{ell}
 a_{11}x^2+2a_{12}xy+a_{22}y^2+2a_{13}x+2a_{23}y=0
 \end{equation}
 \noindent where

\begin{center}
$a_{11}=C^2+4D^2,\quad a_{12}=-(AC+4BD),\quad a_{22}=A^2+4B^2,$ \\
$a_{13}=2D(BC-AD),\quad a_{23}=-2B(BC-AD)$
\end{center}

\noindent and $A=-Re(v), B=-Im(u), C=-Im(v), D=Re(u)$. 
\end{lemma}

\begin{proof}
We express the functions $\xi$ and $\psi$ as
\begin{equation}
\begin{cases}
\xi(\tau)=-v(1-\tau)^2=-v\theta^2=-v\sin^2\varphi,\\
\psi(\tau)=-w(1-(1-\tau)^2)(1-\tau^2)=-w(1-\theta^2)\theta^2=-w\sin^2\varphi\cos^2\varphi,
\end{cases}
\end{equation}
\noindent where $\tau\in[0,1]$, $\theta:=1-\tau\in[0,1]$, and
$\sin\varphi:=\theta$, $\varphi\in[0,\pi/2]$. Then

\begin{center}
$\xi(\tau)\pm2\sqrt{\psi(\tau)}=-v\sin^2\varphi\pm\sqrt{-w}\sin2\varphi.$
\end{center}

Thus the curve $\Gamma\subset \bC$ is  given by the parametrization $ \Gamma(\varphi)=-v \sin^2
\varphi \pm \sqrt{-w}\sin2\varphi $, where $v,w,z\in\Com$ and
$\varphi\in [0,\pi/2]$. Set $w=u^2$, so that $\sqrt{-w}=\iu u$. 
Then $\Gamma$ has the form: 
$$\Gamma(\varphi) =-v \sin^2 \varphi \pm \iu u\sin2\varphi=(-Re(v)-\iu\, Im(v))
\sin^2 \pm \iu(Re(u)+\iu\, Im(u))\sin2\varphi. $$
We, therefore, get the following system for its real and imaginary parts:
\begin{equation}\label{prm}
  \begin{cases}
    x(\varphi)=A \sin^2 \varphi + B \sin2\varphi \\
    y(\varphi)=C \sin^2 \varphi + D \sin2\varphi.
  \end{cases}
\end{equation}
Here $A=-Re(v), B=-Im(u), C=-Im(v), D=Re(u)$ and $\varphi\in
[-\pi/2,\pi/2]$ since $\Gamma$ is $\pi$-periodic.

To show that $\Gamma$ is an ellipse passing through the origin and satisfying (\ref{ell})
  substitute (\ref{prm}) into the expression 
$
a_{11}x^2(\varphi)+2a_{12}x(\varphi)y(\varphi)+a_{22}y^2(\varphi)+
2a_{13}x(\varphi)+2a_{23}y(\varphi)$, where the coefficients $a_{i,j}$ are defined in the statement of Lemma~\ref{ellipse}. 
Simple calculations then show then that the latter expression vanishes identically,  i.e. for all values of $\varphi$.

To prove that   (\ref{ell}) describes a real ellipse (and not some other real affine quadric) consider the determinant

\begin{equation*}
\Delta:=\left|
\begin{array}{c c c}
a_{11} & a_{12} & a_{13} \\
a_{12} & a_{22} & a_{23} \\
a_{13} & a_{23} & 0
\end{array}
\right| =-4(BC-AD)^4.
\end{equation*}
It is well-known that if $\Delta$ is negative then we have a real ellipse ($\Delta>0$ corresponds to an 
imaginary ellipse, i.e. an empty set of solutions). Thus unless $BC-AD=0$ (which describes the situation with all three roots of $Q(z)$ being collinear) then $\Gamma$ is a real ellipse. To find its semiaxes  $a$ and $b$  we calculate the following quantities: 
\begin{equation*}
\delta:=\left|
\begin{array}{c c}
a_{11} & a_{12} \\
a_{12} & a_{22}
\end{array}
\right| =4(BC-AD)^2; \qquad
\iota:=a_{11}+a_{22}=A^2+C^2+4(B^2+C^2).
\end{equation*}
It is known that the roots $\lambda_{1,2}$ of  the characteristic equation
$\lambda^2-\iota\lambda+\delta=0$   are
equal to $2a^2$ and $2b^2$, (in particular, both need to be  positive) where  $a,b$ are the 
semiaxes of the ellipse under consideration. We arrive therefore at
\begin{equation*}
  \begin{cases}
    a=\frac{1}{2} \sqrt{\iota+\sqrt{\iota^2-4\delta}} \\
    b=\frac{1}{2} \sqrt{\iota-\sqrt{\iota^2-4\delta}}
  \end{cases}
\end{equation*}
and
$\sqrt{\iota^2-4\delta}=\sqrt{((A-2D)^2+(C+2B)^2)((A+2D)^2+(C-2B)^2)}$.
For the sake of completeness   the  eccentricity $c$ of our ellipse  can be expressed as

\begin{center}
$c=\sqrt{\frac{-\Delta}{\delta^2}\sqrt{\iota^2-4\delta}}=
\frac{1}{2}\sqrt[4]{((A-2D)^2+(C+2B)^2)((A+2D)^2+(C-2B)^2)}$.
\end{center}

\end{proof}

\begin{lemma}\label{foci}
The foci of the ellipse coincide with the two roots of the polynomial $Q(z)$ different from the origin. 
\end{lemma}

\begin{proof}

The coordinates of the centre $\cc=(x_\cc,y_\cc)$ of our ellipse satisfy:
\begin{equation*}
\left.
\begin{aligned}
a_{11}x_\cc+a_{12}y_\cc+a_{13}=0 \\
a_{12}x_\cc+a_{22}y_\cc+a_{23}=0
\end{aligned}
\right\} \quad \Rightarrow \quad x_\cc=\frac{A}{2} \quad
y_\cc=\frac{C}{2}.
\end{equation*}
Recalling that $Q(z)=z(z^2+vz+w)= z(z^2+vz+u^2)$ we need  to show that the coordinates $(x_f,y_f)$ of the foci $f$ of $\Gamma$ satisfy  the equation:
\begin{equation*}
f=x_f+\iu\, y_f=\frac{-v\pm \sqrt{v^2-4u^2}}{2}.
\end{equation*}
To do this  we express them through $A,B,C,D$. First, we see that
$Re(v^2-4u^2)=A^2+4B^2-C^2-4D^2$. Using the relation:

\begin{center}
$\sqrt{\xi+\iu\, \eta}=\sqrt{\frac{r+\xi}{2}}+\iu\,
\sqrt{\frac{r-\xi}{2}}$,
\end{center}

\noindent where $r=\sqrt{\xi^2+\eta^2}$ we get

\begin{center}
$r=\sqrt{(Re(v^2-4u^2))^2+(Im(v^2-4u^2))^2}=4c^2=$
\end{center}
\begin{center}
$\sqrt{((A-2D)^2+(C+2B)^2) ((A+2D)^2+(C-2B)^2)}$
\end{center}

\noindent and
\begin{equation*}
\begin{aligned}
x_f=\frac{A}{2}\pm\frac{1}{2\sqrt{2}}\sqrt{4c^2+(A^2+4B^2-C^2-4D^2)} \\
y_f=\frac{C}{2}\pm\frac{1}{2\sqrt{2}}\sqrt{4c^2-(A^2+4B^2-C^2-4D^2)}
\end{aligned}
\end{equation*}
Straightforward calculation shows that the centre $\cc$ and the foci $f_1$ and $f_2$ lie on the same line
given by the equation:
\begin{equation}\label{ma}
y=\frac{4c^2-(A^2+4B^2-C^2-4D^2)}{2(AC+4BD)}\left(x-\frac{A}{2}\right)+\frac{C}{2}.
\end{equation}

Finally we check that the spacing between the centre and either focus equals to the eccentricity  $c$ which settles the lemma. This  follows, for example,  from the expression for the coordinates of the intersection points between (\ref{ma}) and the circle $(x-x_\cc)^2+(y-y_\cc)^2=c^2$.
\end{proof}

\section{Proof of Theorem~\ref{th:eq}}

We start with the following integral representation of the required Cauchy trasform. 

\begin{lemma}\label{lm:CT}
The Cauchy transform $\C_0(z)$ of the measure $M_0$ associated with the root of the polynomial $Q(z)=z(z^2+vz+w)$ at the origin is given by
\begin{equation}\label{eq:CT}
\C_0(z)=\int_0^1\frac{d\theta}{\sqrt{(v^2-4w)\theta^4+(2vz+4w)\theta^2+z^2}}.
\end{equation}
\end{lemma} 

\begin{proof} Indeed, recall that the Cauchy transform $\C_{[\alpha_1, \alpha_2]}$Êof the  arcsine measure $\omega_{[\alpha_1, \alpha_2]}$ of the interval $[\alpha_1,\alpha_2]$ equals 
$$\C_{[\alpha_1, \alpha_2]}=\frac{1}{\sqrt{z-\alpha_1)(z-\alpha_2)}}.$$ 
The measure $M_0$ is obtained by the averaging of the family of arcsine measures, namely
$$M_0=\int_0^1
\omega_{\left[\xi(\tau)-2\sqrt{\psi(\tau)},
\xi(\tau)+2\sqrt{\psi(\tau)}\right]}d\tau,$$
where $\xi(\tau)=-v(1-\tau)^2=-v\theta^2$,  $\psi(\tau)=-w(1-(1-\tau)^2)(1-\tau)^2=-w(1-\theta^2)\theta^2$ and $\theta=1-\tau$. Since the Cauchy transform of the average of a family of measures  equals the average of the family of their Cauchy transforms one gets after obvious simplifications:
\begin{equation*}
\begin{aligned}
\C_0(z)=&\int_0^1\frac{d\tau}{(z-(\xi(\tau)-2\sqrt{\psi(\tau)})(z-(\xi(\tau)+2\sqrt{\psi(\tau)})}=\\
=&\int_0^1\frac{d\theta}{\sqrt{(v^2-4w)\theta^4+(2vz+4w)\theta^2+z^2}}.
\end{aligned}
\end{equation*}
\end{proof}Ê

\subsection{Special case}Ê
We first provide the proof of Theorem~\ref{th:eq} for a specific case $Q(z)=z(4z^2-1)$ where the calculations are somewhat simpler and then address the general case. 
By Lemma~\ref{lm:CT} the Cauchy transform $\C_0(z)$ of the measure $M_0$ associate with the root of $Q(z)$ at the origin is then given by the  integral

\begin{equation*}\label{eq:CI}
\C_0(z):=\int_0^1 \frac{d\theta}{\sqrt{\theta^4-\theta^2+z^2}}.
\end{equation*}
We want  to find a differential  equation satisfied by $\C_0(z)$ w.r.t. the variable $z$. Unfortunately, we do not know how to do it directly and our proof requires a number of intricate variable changes and manipulations. We first change  $t=2\theta^2-1$ and consider
\begin{equation}\label{CCII}
\C_0(z)=I_0(s)=\frac{1}{\sqrt{2}}\int_{-1}^1
\frac{dt}{\sqrt{t+1}\sqrt{t^2+s}},
\end{equation}

\noindent where $s:=4z^2-1$. Introduce now a family of functions $I_{\nu}(s)$ indexed by $\nu\ge 0$  and defined by:

\begin{equation*}
I_{\nu}(s):=\frac{1}{\sqrt{2}}\int_{-1}^1
\frac{t^{\nu}dt}{\sqrt{t+1}\sqrt{t^2+s}}.
\end{equation*}

\begin{lemma}\label{lm:relat}
For $\nu\ge 0$ the following three relations are satisfied: 

\bigskip

\begin{equation}\label{rec}
    \frac{\partial
    I_{\nu+2}}{\partial s}=-\frac{1}{2}I_{\nu}-s\frac{\partial
    I_{\nu}}{\partial s},
\end{equation}

\begin{equation}\label{two}
  \frac{\partial}{\partial s}(I_2+I_1)=-\frac{1}{4}I_0+\frac{1}{2\sqrt{1+s}},
\end{equation}

\begin{equation}\label{three}
  \frac{\partial}{\partial
  s}(I_3-I_1)=-\frac{3}{4}I_1-\frac{1}{4}I_0.
\end{equation}

\end{lemma}

\begin{proof} Relation (\ref{rec}) can be proved directly:

\begin{equation*}
\begin{aligned}
    \frac{\partial I_{\nu+2}}{\partial s}
    &=-\frac{1}{2\sqrt{2}}\int_{-1}^1 \frac{t^{\nu+2}dt}{\sqrt{t+1}(t^2+s)^{3/2}}
    \\
    &=-\frac{1}{2\sqrt{2}}\int_{-1}^1
    \frac{(t^{\nu+2}+st^{\nu})dt}{\sqrt{t+1}(t^2+s)^{3/2}}+
    \frac{1}{2\sqrt{2}}\int_{-1}^1
    \frac{st^{\nu}dt}{\sqrt{t+1}(t^2+s)^{3/2}}
    \\
    &=-\frac{1}{2}I_{\nu}-s\frac{\partial I_{\nu}}{\partial s}.
\end{aligned}
\end{equation*}

Relation (\ref{two}) is easy to verify  by integration by parts. Indeed, 

\begin{equation*}
\begin{aligned}
    \frac{\partial}{\partial s}(I_2+I_1)
    &=-\frac{1}{2\sqrt{2}}\int_{-1}^1
    \frac{t+1}{\sqrt{t+1}}\frac{tdt}{(t^2+s)^{3/2}}
    =\frac{1}{2\sqrt{1+s}}-\frac{1}{4}I_0.
\end{aligned}
\end{equation*}

 Similarly, by integration by parts one gets:

\begin{equation*}
\begin{aligned}
    \frac{\partial}{\partial s}(I_3-I_1)
    &=-\frac{1}{2\sqrt{2}}\int_{-1}^1
    \frac{t+1}{\sqrt{t^2-1}}\frac{tdt}{(t^2+s)^{3/2}}
   =-\frac{3}{4}I_1-\frac{1}{4}I_0.
\end{aligned}
\end{equation*}

\end{proof}

Now, we express ${\partial I_2}/{\partial s}$ from \eqref{rec}, substitute 
it in \eqref{two}, and  single out $\partial I_1/\partial s$:

\begin{equation}\label{der1}
    \frac{\partial I_1}{\partial s}=s\frac{\partial I_0}{\partial
    s}+\frac{1}{4}I_0+\frac{1}{2\sqrt{1+s}}.
\end{equation}
Adding \eqref{two} with \eqref{three} and reducing $\partial
I_3/\partial s$, $\partial I_2/\partial s$ with the help of \eqref{rec} we
obtain:

\begin{equation*}
4(1+s)\frac{\partial I_1}{\partial s}=I_0+I_1.
\end{equation*}

\noindent Through   \eqref{der1} we get:

\begin{equation*}
I_1=4s(1+s)\frac{\partial I_0}{\partial s}+s I_0+2\sqrt{s+1}.
\end{equation*}

\noindent Differentiating both sides of the latter relation w.r.t. 
 $s$ and using
\eqref{der1} again we obtain the required linear non-homogeneous  differential equation satisfied by $I_0(s)$:

\begin{equation}\label{eq}
16s(1+s)\frac{\partial^2 I_0}{\partial s^2}+16(1+2s)\frac{\partial
I_0}{\partial s}+3I_0=-\frac{2}{\sqrt{1+s}}.
\end{equation}

In order to recover the required equation \eqref{eq:Heun} for $\C_0(z)$ we have to change $s$ back to $z$. Using straightforward  relations
$$
\frac{\partial I_0}{\partial s}=\frac{1}{8z}\frac{\partial \C_0}{\partial z}\text{    and    }
\frac{\partial^2 I_0}{\partial s^2}=\frac{1}{64z^3}\left(z\frac{\partial^2 \C_0}{\partial z^2}-\frac{\partial \C_0}{\partial z}\right)
$$
we obtain after some obvious simplifications the equation: 
$$z(4z^2-1)\frac{\partial^2 \C_0}{\partial z^2}+(12z^2-1)\frac{\partial \C_0}{\partial z}+3z\C_0(z)+1=0$$
which coincides with \eqref{eq:Heun} for $Q(z)=z(4z^2-1)$. Thus our special case of Theorem~\ref{th:eq} is settled. 

Notice also that  \eqref{eq} can be solved explicitly. The general solution of
the corresponding linear homogeneous equation is an arbitrary  linear combination of a complete elliptic
integral of the first kind $y_1(s)$ and of an  associated Legendre function of
the second kind $y_2(s)$ given by:

\begin{equation*}
\begin{cases}
y_1(s)=&
\frac{2}{\pi\sqrt[4]{1+s}}\mathbb{K}\left(\frac{\sqrt{1+s}-1}{2\sqrt{1+s}}\right)
\\
y_2(s)=& \mathbb{Q}_{-1/4}(1+2s), 
\end{cases}
\end{equation*}
here $\mathbb{K}(x)$ and $\mathbb{Q}(x)$ are the complete elliptic integral and the associated Legendre function of the second kind respective.
\noindent The general solution to \eqref{eq} depends on  two arbitrary
constants $C_1, C_2$ and is given by:

\begin{equation*}
I_0=C_1y_2+C_2y_2+y_2\int y_1\frac{g}{f_2}\frac{ds}{W}-y_1\int
y_2\frac{g}{f_2}\frac{ds}{W},
\end{equation*}

\noindent where
$g(s)=-\frac{2}{\sqrt{1+s}}, \quad f_2(s)=16s(1+s), \quad
W(s)=y_1(s)y'_2(s)-y_2(s)y'_1(s).$
%
However, we need its particular solution and thus  have to determine the corresponding 
particular values of $C_1, C_2$. (To find them  we  evaluated the
integral \eqref{CCII} for two different values of $s$. Moreover, analyzing the polynomial $(t+1)(t^2+s)=t^3+t^2+st+s$, we observed  that it is positive on $[-1,1]$ for $s>0$ and that  \eqref{CCII} is divergent when $s=0$.) 

The next figure compares  the appropriate solution of  \eqref{eq} giving $I_0(s)$ with the values of $I_0(s)$ calculated numerically using the integral \eqref{CCII}  for a number of values of $s$ (which are shown by dots below).
\begin{figure}[h!]\label{fig3}
\epsfig{file=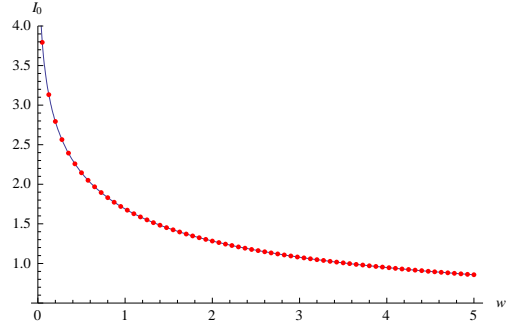, width=6.5cm} \caption{The graph of $I_0(s)$ obtained from \eqref{eq} and its numerical values obtained from \eqref{CCII}.}
\end{figure}
%

\subsection{General case}

The scheme of this proof is exactly the same as in the above special case but  calculations are somewhat messier. 
Assuming that $v^2-4w\neq0$ we need to find a differential equation satisfied by the integral \eqref{eq:CT}.
 We change variables as follows:
\begin{equation*}
s=-16w\frac{z^2+vz+w}{(v^2-4w)^2}, \qquad u=v\frac{v+2z}{v^2-4w},
\qquad a=v^2-4w
\end{equation*}
and denote $I_0(s,u,a)=\C_0(z)$. (Here as above we assume that $v$ and $w$ are some fixed complex numbers.)  It also helps  to change the 
variable $\theta$ in \eqref{eq:CT} by using $2\theta^2=t+1$ and then we finally get
\begin{equation*}
\C_0(z)=I_0(s)=\frac{1}{\sqrt{2a}}\int_{-1}^1\frac{dt}{\sqrt{t+1}\sqrt{(t+u)^2+s}}.
\end{equation*}
As above we introduce a family of functions $I_{\nu}(s),\;s\ge0$ given by the formula:
\begin{equation*}
I_\nu(s):=\frac{1}{\sqrt{2a}}\int_{-1}^1\frac{(t+u)^\nu
dt}{\sqrt{t+1}\sqrt{(t+u)^2+s}}. 
\end{equation*}
Analogously to Lemma~\ref{lm:relat} one can prove the next statement.  

\begin{lemma} The following relations are valid for $I_{\nu}(s),\;s\ge0$:
\begin{equation}\label{srec}
    \frac{\partial
    I_{\nu+2}}{\partial s}=-\frac{1}{2}I_{\nu}-s\frac{\partial
    I_{\nu}}{\partial s},
\end{equation}

\begin{equation}\label{stwo}
   \frac{\partial}{\partial s}(I_2+I_1)=u\frac{\partial I_1}{\partial s}
  -\frac{1}{4}I_0+\frac{1}{2\sqrt{2}\sqrt{(u+1)^2+s}},
\end{equation}

\begin{equation}\label{sthree}
  \frac{\partial}{\partial
  s}(I_3-I_1)=(u^2-2u)\frac{\partial I_1}{\partial s}-\frac{3}{4}I_1+
  \frac{u-1}{4}I_0+\frac{u}{\sqrt{2}\sqrt{(u+1)^2+s}}.
\end{equation}
\end{lemma}

Now, we use \eqref{srec} for expressing ${\partial I_2}/{\partial
s}$ and then we single out ${\partial I_1}/{\partial s}$ from \eqref{stwo}:
\begin{equation}\label{sder1}
\frac{\partial I_1}{\partial s}=\frac{s}{1-u}\frac{\partial
I_0}{\partial
s}+\frac{I_0}{4(1-u)}+\frac{1}{2\sqrt{a}(1-u)\sqrt{(u+1)^2+s}}.
\end{equation}
Adding \eqref{stwo} and \eqref{sthree}, employing \eqref{srec} again,
and using \eqref{sder1} we get the relation:
\begin{equation}\label{sI1}
  (u-1)I_1=-4s(s+(u-1)^2)\frac{\partial I_0}{\partial
  s}-sI_0-\frac{2(s-u^2+1)}{\sqrt{a}\sqrt{(u+1)^2+s}}.
\end{equation}
Eventually, taking the derivative of the both sides of the latter equation w.r.t $s$ and using
\eqref{sder1} again we finally get a linear differential equation in the variable $s$ satisfied by $I_0(s)$:
\begin{equation}\label{seq}
16s(s+(u-1)^2)\frac{\partial^2 I_0}{\partial
  s^2}+16(2s+(u-1)^2)\frac{\partial I_0}{\partial
  s}+3I_0+\frac{2}{\sqrt{2}}\frac{s+(u+1)(5u+1)}{\sqrt{(u+1)^2+s}}=0.
\end{equation}

In order to get an equation for $\C_0(z)$ w.r.t. the variable $z$, we use:
\begin{equation*}
\frac{\partial \C_0}{\partial z}=\frac{\partial s}{\partial
z}\frac{\partial I_0}{\partial s}+\frac{\partial u}{\partial
z}\frac{\partial I_0}{\partial u}
\end{equation*}
and
\begin{equation*}
  \frac{\partial^2 \C_0}{\partial z^2}=\frac{\partial^2 s}{\partial z^2}
  \frac{\partial I_0}{\partial s}+\left(\frac{\partial s}{\partial
  z}\right)^2\frac{\partial^2 I_0}{\partial s^2}+2\frac{\partial s}{\partial z}
  \frac{\partial u}{\partial z}\frac{\partial^2 I_0}{\partial s \partial
  u}+\left(\frac{\partial u}{\partial
  z}\right)^2\frac{\partial^2 I_0}{\partial u^2}.
\end{equation*}
With the help  of $\frac{\partial I_0}{\partial u}=2I_1$
and \eqref{sI1} we obtain
\begin{equation*}
\frac{\partial \C_0}{\partial z}=\left(\frac{\partial s}{\partial
z}+2\frac{\partial u}{\partial
  z}\frac{s}{1-u}\right)\frac{\partial I_0}{\partial s}+2\frac{\partial u}{\partial
  z}\frac{I_0}{4(1-u)}+\frac{\partial u}{\partial
  z}\frac{1}{\sqrt{a}(1-u)\sqrt{(u+1)^2+s}}.
\end{equation*}
Now, we get
\begin{equation}\label{J0z}
\frac{\partial I_0}{\partial
s}=\frac{(v^2-4w)(vz+2w)}{16wz}\frac{\partial \C_0}{\partial
z}+\frac{v(v^2-4w)}{32wz}\C_0+\frac{v(v^2-4w)}{32wz(v+z)}.
\end{equation}

Further, we use
\begin{equation*}
\frac{\partial^2 I_0}{\partial u^2}=-4\frac{\partial I_0}{\partial
s}-4s\frac{\partial^2 I_0}{\partial s^2}
\end{equation*}
and
\begin{equation*}
\begin{aligned}
\frac{\partial^2 I_0}{\partial s \partial
u}=&2(u-1)\frac{\partial^2 I_0}{\partial
s^2}+\frac{3s+4(u-1)^2}{2s(u-1)}\frac{\partial I_0}{\partial
s}+
\frac{3I_0}{8s(u-1)}+\frac{3s+5u^2+6u+1}{4\sqrt{a}(u-1)s(s+(u-1)^2)^{3/2}}.
\end{aligned}
\end{equation*}
We can now express ${\partial^2 \C_0}/{\partial z^2}$ through
${\partial^2 I_0}/{\partial s^2}$, ${\partial
I_0}/{\partial s}$, and $I_0$ as follows:
\begin{equation*}
\begin{aligned}
\frac{\partial^2 \C_0}{\partial z^2}=&\left(\left(\frac{\partial
s}{\partial z}\right)^2+4(u-1)\left(\frac{\partial s}{\partial
z}\right)\left(\frac{\partial u}{\partial
z}\right)-4s\left(\frac{\partial u}{\partial
z}\right)^2\right)\frac{\partial^2 I_0}{\partial s^2}+
\\
 &+\left(\frac{\partial^2 s}{\partial z^2}+\left(\frac{\partial s}{\partial
 z}\right)\left(\frac{\partial u}{\partial
 z}\right)\frac{3s+4(u-1)^2}{s(u-1)}-4\left(\frac{\partial u}{\partial
 z}\right)^2\right)\frac{\partial I_0}{\partial s}+
\\
 &+\left(\frac{\partial s}{\partial
 z}\right)\left(\frac{\partial u}{\partial
 z}\right)\frac{3I_0}{4s(u-1)}+\left(\frac{\partial s}{\partial
 z}\right)\left(\frac{\partial u}{\partial
 z}\right)\frac{3s+5u^2+6u+1}{4\sqrt{a}(u-1)s(s+(u-1)^2)^{3/2}}.
\end{aligned}
\end{equation*}
This leads to:

\begin{equation*}
\begin{aligned}
\frac{\partial^2 \C_0}{\partial
z^2}=&-\frac{256wz^2}{(v^2-4w)^3}\frac{\partial^2 I_0}{\partial
s^2}-
\\
 &-16\frac{4w^2(w+z^2)+4vwz(w+2z^2)+v^2w(w+5z^2)+v^3(2wz-z^3)}
{(v^2-4w)^2(2w+vz)(w+z(v+z))}\frac{\partial I_0}{\partial s}+
\\
 &+\frac{3v(v+2z)}{4(2w+vz)(w+z(v+z))}I_0+
\\
 &+\frac{v(v+2z)(3v^4+8v^3z-24vwz-4w(2w+3z^2)
 +v^2(5z^2-8w))}{4(v^2-4w)(v+z)^3(2w+vz)(w+z(v+z))}.
\end{aligned}
\end{equation*}
From the latter equation and \eqref{J0z} we finally get:
\begin{equation*}
\begin{aligned}
\frac{\partial^2 I_0}{\partial
s^2}=&-\frac{(v^2-4w)^3}{256wz^2}\frac{\partial^2 \C_0}{\partial
z^2}- \frac{(v^2-4w)^2 c_1}
 {256w^2z^3(w+z(v+z))}\frac{\partial \C_0}{\partial z}-
 \\
 &-\frac{v(v^2-4w)^2c_2}{1024wz^2(w+z(v+z))}\C_0-
\frac{v(v^2-4w)^2c_3}
 {1024w^2z^3(v+z)^3(w+z(v+z))},
\end{aligned}
\end{equation*}
where
\begin{equation*}
\begin{aligned}
c_1=&4w^2(w+z^2)+4vwz(w+2z^2)+v^2w(w+5z^2)+v^3(2wz-z^3),
\\
c_2=&8vwz+v^2(w-2z^2)+4w(w+4z^2)),
\\
c_3=&v^4(w-2z^2)+12vwz(w+3z^2)+4wz^2(3w+4z^2)+v^3(8wz-4z^3)+
\\
 &+v^2(4w^2+27wz^2-2z^4).
\end{aligned}
\end{equation*}

Plugging these formulae into \eqref{seq} we arrive  at:
\begin{equation*}
4z(z^2+vz+w)\C_0^{''}(z)+4(3z^2+2vz+w)\C_0^{'}(z)+(3z+v)\C_0(z)+1=0,
\end{equation*}
which can be equivalently expressed as
%
\begin{center}
\fcolorbox{LightGreen1}{white}{ $
Q(z)\C_0^{''}(z)+Q^{'}(z)\C_0^{'}(z)+Q^{''}(z)\C_0(z)/8+Q^{'''}(z)/24=0,
$\,\rule{0ex}{2.2ex}}
\end{center}
with $Q(z)=4z(z^2+vz+w)$. (Notice that the multiplication of $Q(z)$ by a non-vanishing constant is irrelevant in our considerations.) \qed

\section{Final remarks}

It is very tempting to extend the methods and results of the present paper to the case of the 'generalized' Heun equations which are of the form 
$$\left\{Q_{k+1}(z)\frac{d^k}{dz^k}+Q_{k}(z)\frac{d^k}{dz^{k-1}}+...+Q_2(z)\frac{d}{dz}+V(z)\right\}S(z)=0,$$
where $\deg Q_{k+1}(z)=k+1$ and $\deg Q_i(z)\le i$ for $i=2,3,...,k$. As in the introduction for each positive (and sufficiently large) integer $n$ there exist $n+1$ polynomials $V(z)$ counted with appropriate multiplicities such that for each of these $V(z)$ the above equation has a polynomial solution $S(z)$  of degree $n$. Thus one can define the corresponding spectral polynomials and study the asymptotics of their root-counting measures. Large scale  numerical experiments support the following. 

\begin{figure}[!htb]
\centerline{\hbox{\epsfysize=4.5cm\epsfbox{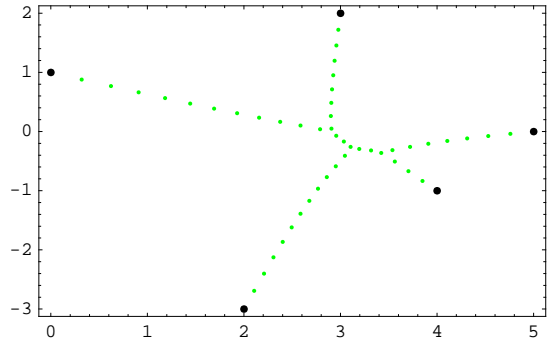}}}
\caption{The measure $\mu$ for the operator $Q(z)\frac{d^4}{dz^4}$ with $Q(z)=(z-5)(z-I)(z-4+I)(z-2+3I)(z-3-2I)$. }
\label{fig4}
\end{figure}

\begin{conjecture}
For any 'generalized' Heun equation  the sequence $\{\mu_n\}$ of the root-counting measures of its spectral polynomials converges  to a probability measure $\mu$ supported on a curvilinear planar tree  located inside $Conv_{Q_{k+1}}$ and whose leaves (i.e. vertices of valency $1$) is the set of  all roots of $Q_{k+1}(z)$, see Fig.~\ref{fig4}. Moreover, the limiting measure $\mu$ depends only on $Q_{k+1}(z)$, i.e. is {\em independent} of the other coefficient of the equation. 
\end{conjecture}

We finish our paper with the following problem.

\begin{pr} Under the assumption that the latter conjecture holds (which is very likely) is it true that the Cauchy transform $\C_\mu$ of the limiting root-counting measure $\mu$ satisfies a linear ode of the form:
$$Q_{k+1}(z)\C_\mu^{(k)}(z)+a_1Q'_{k+1}(z)\C_\mu^{(k-1)}(z)+a_1Q_{k+1}^{\prime\prime}(z)\C_\mu^{(k-2)}(z)+...+a_{k+1}Q_{k+1}^{(k+1)}(z)=0,$$
where $a_1,...,a_k$ are some universal constants, i.e. independent of  $Q_{k+1}(z)$ (but maybe dependent on the order $k$ of the operator).

\end{pr}

\end{document}